\documentclass[conference]{IEEEtran}
\IEEEoverridecommandlockouts
\usepackage{url}
\usepackage{cite}
\usepackage{amsmath,amssymb,amsfonts}
\usepackage{algorithmic}
\usepackage{graphicx}
\usepackage{textcomp}
\usepackage{xspace}
\usepackage{xcolor}
\usepackage{array} \usepackage{multirow}
\usepackage{tikz}
\usepackage[hidelinks]{hyperref} \usepackage{xurl}

\usepackage{hyperref}
\usepackage{tablefootnote}
\usepackage{subcaption}

\newcommand{\toolname}{{{{cyberaCTIve}}}}

\def\BibTeX{{\rm B\kern-.05em{\sc i\kern-.025em b}\kern-.08em
    T\kern-.1667em\lower.7ex\hbox{E}\kern-.125emX}}

\begin{document}

\title{\toolname: a STIX-based Tool for Cyber Threat Intelligence in Complex Models}

\author{
\IEEEauthorblockN{Ricardo M. Czekster\IEEEauthorrefmark{1}, Roberto Metere\IEEEauthorrefmark{2}\IEEEauthorrefmark{3}, Charles Morisset\IEEEauthorrefmark{2}}
\IEEEauthorblockA{\IEEEauthorrefmark{1}School of Informatics and Digital Engineering, Aston University, Birmingham, United Kingdom\\r.meloczekster@aston.ac.uk}
\IEEEauthorblockA{\IEEEauthorrefmark{2}School of Computing, Newcastle University, Newcastle upon Tyne, United Kingdom\\\{roberto.metere, charles.morisset\}@ncl.ac.uk}
\IEEEauthorblockA{\IEEEauthorrefmark{3}The Alan Turing Institute, London, United Kingdom}
}

\maketitle
\thispagestyle{plain}
\pagestyle{plain}

\begin{abstract}
Cyber threat intelligence (CTI) is practical real-world information that is collected with the purpose of assessing threats in cyber-physical systems (CPS).
A practical notation for sharing CTI is STIX.
STIX offers facilities to create, visualise and share models; however, even a moderately simple project can be represented in STIX as a quite complex graph, suggesting to spread CTI across multiple simpler sub-projects.
Our tool aims to enhance the STIX-based modelling task in contexts when such simplifications are infeasible.
Examples can be the microgrid and, more in general, the smart grid.
\end{abstract}


\section{Introduction}
\label{sec:intro}
Criminal activity has increased as workers shifted towards virtual alternatives to perform their tasks~\cite{buil2021cybercrime,lallie2021cyber}.
These malicious actors exploit vulnerabilities that stretch resources to the limit for financial gains, impair trust, recognition, or data theft, among other reasons. 

In many cases, performing remote tasks is the only alternative, and remote control features translate to a growth of interconnected cyber-physical systems (CPS).
One example of a CPS is the smart grid where power managers implement dynamic load responses for different energy profiles to meet supply-demand~\cite{alotaibi2020comprehensive}.

What makes hard to deter cyber-criminals and deviant behaviour is the overwhelming time needed to understand malicious occurrences, complex timelines, trusting issues related to sources, to then act on protecting the infrastructure.
Generally, such information is collected under the term {\em cyber threat intelligence} (CTI)~\cite{tounsi2019cyber}\footnote{Also termed {\em information security threat intelligence}, or ISTI; however, we shall use CTI throughout this work.}.
As a system gets eventually compromised and then quickly reported, other organisations can be made aware and can prepare themselves to respond to similar potential threats.
This exchange of CTI highlights the importance of effectively describe, report, and share it by using standardised and structured formats comprehensible by analysts across diverse disciplines.

This work extends on the standard format used by cyber-security analysts called STIX\textsuperscript{TM} (Structured Threat Information eXpression)~\cite{barnum2012standardizing}.
STIX allows to model, visualise and share CTI models that collect valuable insights and contextual intelligence to establish richer and wider cyber-attack narratives to understand ongoing incursions.
These data are valuable tools helping security officers take prompt actions to thwart cyber-attacks or protect systems against further advances.
However, the main difficulty of STIX, despite having a very thorough documentation, is due to its inherent complexity when creating or updating models given the sheer amount of parameters one must consider.

In this paper we are interested in helping cyber-security officers to address incorporating CTI related data into smart buildings.
Our contribution offers a partial solution that takes into account all the problems outlined above.
In particular, we propose a tool for modelling structured cyber-attacks called \toolname{} (Cyber-Attack CTI Virtual Environment).
Our aim is to ease modelling and structuring cyber-attack progression as new pieces of evidence emerge through different sources, as well as to simplify sharing and analysing CTI.

Furthermore, we implemented a log of all events that analysts created and order them by the timestamp.
This extension allows for a {\em timeline analysis}, where analysis could inspect the {\em chain of events} and then try to reason on ongoing cyber-attacks.
This can help finding patterns that can apply to corresponding real-world data audits and speed up forensic analysis.
For example, they may quickly review other models and add more contextual data, clarify their position, or deciding to employ mitigating pro-active measures to thwart attacks.
Overall, we simplify STIX-based modelling effort and ensure high cyber-attack expressiveness to help thwart attacks and disseminate relevant contexts for peers to react faster.

\section{Cyber Threat Intelligence}\label{sec:cti}
Data, information, knowledge, intelligence and wisdom are interrelated concepts that play a relevant role for an effective system analysis~\cite{ackoff1989data}.
The terminology \textit{Threat Intelligence} itself originates from military literature that was later broadly adopted by other stakeholders.
Industrial partners, more specifically the Gartner group, acknowledges the following assertion: \textit{``Threat intelligence is evidence-based knowledge, including context, mechanisms, indicators, implications and actionable advice, about an existing or emerging menace or hazard to assets that can be used to inform decisions regarding the subject’s response to that menace or hazard.''}\footnote{\url{https://www.gartner.com/en/documents/2487216/definition-threat-intelligence}.}.

Recent advances introduced the term cyber threat intelligence (CTI)~\cite{tounsi2018survey,tounsi2019cyber} to encompass information technologies.
CTI is thus a way of anticipating attacks or malicious incursions in networks deemed crucial for maintaining business operations.
It helps stakeholders to devise measures to complicate, prevent, or thwart adversarial actions in networks where managers require control and security.
Major individuals interested in employing CTI are network administrators, security officers, managers and decision makers, incident responders, and infrastructure operators.

CTI holds value to enterprises wishing to enlarge the analysis scope when considering cyber-attacks.
We mention the SANS Institute\footnote{SANS is a US institute and the acronym stands for SysAdmin, Audit, Network, and Security.} report tackling how it is employed by organisations~\cite{brown2019evolution}.
They noticed an increase in interest over the years, however, stakeholders comment on the need of expanding use cases to enhance how to understand the CTI benefit and security posture gains.
The report also discusses on the need for improve report automation and ways of enlarging adopting by government-sponsored groups, private sector, and industry-focused groups, to name a few of their findings.

Examples of data sources used in intelligence gathering~\cite{brown2019evolution} stretches technical, human, and internal domains, and it could be both structured and unstructured~\cite{tounsi2019cyber,pokorny2019threat}.
Quality of CTI-based feeds is a topic of wide interest~\cite{schaberreiter2019quantitative,griffioen2020quality} in attempts to determine best data sources with high-quality curated cyber-security withholding crucial elements to make decisions.
Tundis et al. (2019)~\cite{tundis2020automated}, for instance, investigated automated assessment of sources and computed a relevance score index to reduce the time needed to verify gathered intelligence.

Security officers and CI managers should assess and evaluate data available in open or public CTI feeds, data from security vendors, industry reports on vulnerabilities (zero day, etc.), open source intelligence (OSINT) reports\footnote{By `Open', in this community, it refers to publicly available data on security.}, security data extracted from IDS or firewall, data from the security, information, and event management (SIEM) platform, incident response systems, and network traffic and flow logs, to mention a few.
Ramsdale et al. (2020)~\cite{ramsdale2020comparative} conducted a comparative analysis of threat intelligence sources, highlighting structured standards such as STIX~\cite{barnum2012standardizing}, Trusted Automated Exchange of Indicator Information (TAXII\textsuperscript{TM})~\cite{connolly2014trusted}, and Cyber Observable eXpression (CybOX\textsuperscript{TM})~\cite{barnum2012cybox,casey2015leveraging}.

CTI, however, could lack on timeliness, i.e., by the time crucial information is processed and disseminated over peers, it could be already outdated.
On the one hand, the use of CTI feeds and sharing features undoubtedly helps deter malicious activities.
On the other hand, advertising exploited vulnerabilities could be an incentive for triggering cyber-attacks as malicious actors will be aware of impending threats.
If security teams are not responsive in due course, their omissions could lead to irreversible damages, data exfiltration, ransomware, or financial loss.
Another drawback of CTI concerns lack of reporting, or under-reporting malicious activities.
The cause varies from case to case but is usually related to companies being afraid of losing user trust on their systems, negative publicity, competitive disadvantage, lawsuits, or liabilities due to lack of preparedness.

One could employ CTI against advanced persistent threats~\cite{sood2012targeted,alshamrani2019survey} and threat actors attempting long term cyber-physical malicious incursions.
To that effect, authors have combined risk assessment with threat modelling to better understand these kinds of cyber-attacks~\cite{tatam2021review}.
Cyber-security managers could devise strategies on how to conduct effective data poisoning attacks over long periods of time to skew algorithms and impair flexible control.
In terms of modelling and representing CTI related data, Takahashi et al. (2012)~\cite{takahashi2012iodef} has conducted research on structured cyber-security reporting in incident object description exchange format (IODEF).
Industrial counterparts have also taken an interest in CTI, for instance, the Mandiant organisation has published a white paper with regards to open indicators of compromise (OpenIOC)~\cite{mandiant2014open} whereas the real-time inter-network defence (RID)~\cite{moriarty2010real}, a request for comments (RFC) number $6545$\footnote{\url{https://www.hjp.at/doc/rfc/rfc6545.html}.}, outlined a proactive method to help sharing incident data sanctioned by the Internet Engineering Task Force (IETF).

Wagner et al. (2019)~\cite{wagner2019cyber} discussed CTI sharing in a survey, explaining basic concepts and future research directions.
Rudman and Irwin (2016)~\cite{rudman2016dridex} devised a tool that used samples from the peer-to-peer (P2P) malware \textit{Dridex}\footnote{United States Computer Emergency Readiness Team (US-CERT). (2015, October) Alert (TA15-286A) Dridex P2P Malware.} for generating IoC in an automated fashion.
Mavroeidis and Bromander (2017)~\cite{mavroeidis2017cyber} discussed ontologies, sharing standards, and taxonomies for tackling CTI.
The work compares different methodologies and existing model's expressiveness for use by security officers.

One cannot neglect the importance and impact of incorporating machine learning (ML) or artificial intelligence (AI) algorithms and techniques into cyber-security applications.
Modern AI employs techniques combined with advanced statistics to help in situations where one requires automated decisions.
AI excels in speed in accuracy in contrast to time-consuming manual processing.
Analysts use it in applications occurring in well bounded problems where the solution and the method to extract insight is within the data.
For problems requiring additional context AI suffers to find good solutions, at least in current research prognostics.
Merging AI with cyber-security aims to automate decisions for inferring `under attack' statuses or confirming malicious activities.

Kaloudi and Li (2020)~\cite{kaloudi2020ai} discussed how AI and ML may help cyber-attackers devise and exploit vulnerabilities in systems, proposing a framework to tackle this new malicious incursion opportunity.
The idea and scope of the work is to survey the literature for highly sophisticated attacks and novel threats to enhance preparedness.
Truong et al. (2020)~\cite{truong2020artificial} commented on the uses and limitations of AI/ML in cyber-security, highlighting prospects and challenges whereas Iqbal and Anwar (2020)~\cite{iqbal2020scerm} proposed a system for providing automated CTI elements to stakeholders.

\section{Related work}
\label{sec:related}

There are significant initiatives to deal with these malicious incursions.
MITRE, a US based organisation, has developed the adversarial tactics, techniques, and common knowledge (ATT\&CK\textregistered) framework~\cite{strom2018mitre} and defined `matrices' (namely enterprise and mobile domains) to help stakeholders understand the tactics, techniques, and procedures (TTP) deployed by attackers.
MITRE has also introduced a similar initiative for applying ATT\&CK to industrial control systems (ICS) called ATT\&CK for ICS~\cite{alexander2020mitre}, due to observed particularities in these systems.
The ATT\&CK framework superseded the common attack pattern enumeration and classification (CAPEC)~\cite{roberts2021cyber}, and we witness academic and industrial partners engaging with reporting efforts to mitigate cyber-attacks.

Administrators may use ATT\&CK to search for evidences (in their own networks) of malicious incursions that happened elsewhere, as well as the related TTPs behind attacks.
The tactics explain `why?' they needed to do the attack whereas the `how?' details the steps undertaken to reach that objective.
Procedures glue everything together giving context to the attack so the analyst may understand the motivations behind the actions and the ways they could establish footholds and persistence.

The richness and descriptive nature of the malicious encounters help analysts investigate most significant details that could help preventing next cyber-attacks.
Unfortunately, a massive number of incursions remains not reported or under-reported.
Reasons vary, and are mainly due to financial consequences of disclosing these data as well as liability or shame in recognising that the company was victim of an attack.

The framework is a valuable resource to help security officers to counteract cyber-attacks with threat-informed defences.
ATT\&CK differs from classic Cyber Kill Chain\textregistered~(CKC)~\cite{hutchins2011intelligence} pioneered by Lockheed Martin~\cite{martin2015seven} in the sense that it identifies and maps adversarial actions that could happen without any order.
Kwon et al. (2020)~\cite{kwon2020cyber} has created a method for translating ATT\&CK matrix threats directly into NIST's cybersecurity framework.
This clearly shows the need to cross-reference models altogether helping cyber-security experts in their tasks.

The community effort to establish public databases of software vulnerability relates information to the Common Vulnerability Scoring System (CVSS)~\cite{mell2006common}, used along with the US National Vulnerability Database (NVD)\footnote{National Vulnerability Database. \url{https://nvd.nist.gov/}.}.
The NVD uses CVSS to track, score, document, and describe details about discovered vulnerabilities reported by industrial partners and individuals.
Computing a scoring system that is vouched by the cyber-security expert community is invaluable for practitioners, since the numeric index provides a notion on severity and the vulnerability's impact on the infrastructure. 
We mention also that the MITRE Corporation, in cooperation with the NIST and the NVD, maintains the Common Vulnerabilities and Exposures (CVE) database\footnote{Common Vulnerabilities and Exposures. \url{https://cve.mitre.org/}.}.
We have compiled a non-comprehensive list of databases and repositories worth of note in Appendix~\ref{app:vuln-repos}.

In a quantitative venue applied to cyber-security throughout the years, research has tackled metrics indices for modelling incidents in CI~\cite{pendleton2016survey,ramos2017model}.
There is interest in this community concerning the list of key quantitative metrics and indicators of attacks being perpetrated by malicious actors so security officers may improve analysis.
For example, Ramos et al. (2017)~\cite{ramos2017model} has surveyed research on metrics whereas Abraham and Nair (2014)~\cite{abraham2014cyber} have addressed the use of quantitative security measures to aid security engineers.
They characterised four security classes of metrics: i) Core, e.g., CVSS, total vulnerability measure (TVM), or Langweg metric (LM); ii) Probability based; iii) Structural, such as shortest path, Number of paths, or mean of path lengths; or iv) Time based, for instance, mean time to recovery (MTTR), mean time to first failure (MTFF), or mean time to breach (MTTB).
In close relation to cyber-security metrics, one could combine with indicators of compromise (IoC)~\cite{tounsi2019cyber}, for example, IP address lifetime, or malware signatures.

\section{\toolname: a tool for assisting CTI}
\label{sec:results}

We propose here a tool to extend STIX-based modelling in a single analysis environment called \toolname, a short name for cyber-attack CTI virtual environment.
It aims to help cyber-security officers in large cyber-physical systems to model, understand, and share CTI-related models among peers to quickly thwart cyber-attacks before they propagate and harm other parts of networks.

A straightforward front-end to STIX databases is not a novel concept, i.e., we are aware of STIX 2.1 Drag and Drop Modeler\footnote{\url{https://github.com/STIX-Modeler/UI}. The tool uses auxiliary libraries such as React, MobX, and Webpack.} where authors have used JavaScript to create models using the standard.
One of its usability issues is that it does not directly handle projects, in the sense that saving and loading projects must be done manually and any accidental refresh of the page would irrecoverably erase all the current project.
What differentiates our tool is the ability of offering a management interface of models/objects and enable users to form analysis groups as they may interact only with models pertaining the same profile.
On top of these differences, our tool uses auxiliary JSON files to cope with the current and also future STIX versions and customised features such as showing {\em timelines} of selected events.

As cyber-attacks progress in the infrastructure over time, security analysts must start producing valuable insights based on their observations about simple anomalies or odd user behaviours.
One could document these abnormalities using spreadsheets, internal memos, or using specific applications, however, there is a need to employ structured modelling of cyber-security events happening in networks.
In this regard, STIX is a standard and modelling language defining modelling primitives that map most likely TTPs that could occur.
Examples of the entities present in the framework are STIX Domain Objects (SDO), STIX Relationship Objects (SRO), and STIX Cyber-observable Objects (SCO).
These elements present modellers with a small set of essential parameters for unambiguously depicting any malicious incursion.
It is an invaluable tool for helping analysts understand and reason about abnormalities that could potentially be classified as active attacks.

STIX uses SCOs for characterising host-based and network-based information. 
They are used by various SDOs to provide supporting context. 
The `Observed data' SDO, for example, indicates that the raw data was observed at a particular time.
A STIX model can be visualised as a graph whose nodes are SDOs and SCOs and whose edges linking nodes are SROs.
These STIX entities are essential to model TTPs, however, one must understand many model-related intricacies before even starting modelling them within the framework.
Users must cope with a large parameter space required by the entities to create usable and actionable high-level models.
\toolname{} helps users create STIX models and add objects in a simple manner.
The system is web-based and allows (personal) projects to be easily stored and retrieved for analysis at any time.

A key feature we have implemented in the tool is the idea of a {\em timeline} that complements previous insights with new information and situations that are unfolding.
We notice that the main difficulty when modelling cyber-attacks is due to low understanding and differentiating stressful situations that are overwhelming resources from active malicious activities.
And that is even harder if APT are to consider in a comprehensive cyber-security assessment.
So, here we propose a simplification on the way of modelling cyber-attacks, reasoning, and abstracting circumstances to enrich analysis and evaluation.

\subsection{Modelling cyber-security incidents with STIX}
We chose to use and represent attacks with the structured notation offered by STIX's documentation~\cite{jordan2019stix}.
The language and format specification defines $18$ SDOs, two SROs, and $36$ SCOs (among types and sub-types).
The SDOs are: 1) Attack Pattern; 2) Campaign; 3) Course of action; 4) Grouping; 5) Identity; 6) Indicator; 7) Infrastructure; 8) Intrusion set; 9) Location; 10) Malware; 11) Malware analysis; 12) Note; 13) Observed data; 14) Opinion; 15) Report; 16) Threat actor; 17) Tool; and 18) Vulnerability.
The SROs are 1) Relationship; and 2) Sighting, whereas SCOs are for example, \texttt{files}, \texttt{network-traffic}, and \texttt{directory} types, to mention a few.
The latter represent any piece of data that is important to convey context to cyber-attacks.

At first glance, it seems rather simple to convey even sophisticated cyber-attacks using STIX.
However, there are literature discussing problems even for experienced STIX modellers when representing simple attacks.
This is the main drive to our research and proposition here, as we aim to simplify the overall process of effectively modelling with STIX, while maintaining its core functionality.

\begin{figure}[!htb]
\includegraphics[width=\columnwidth]{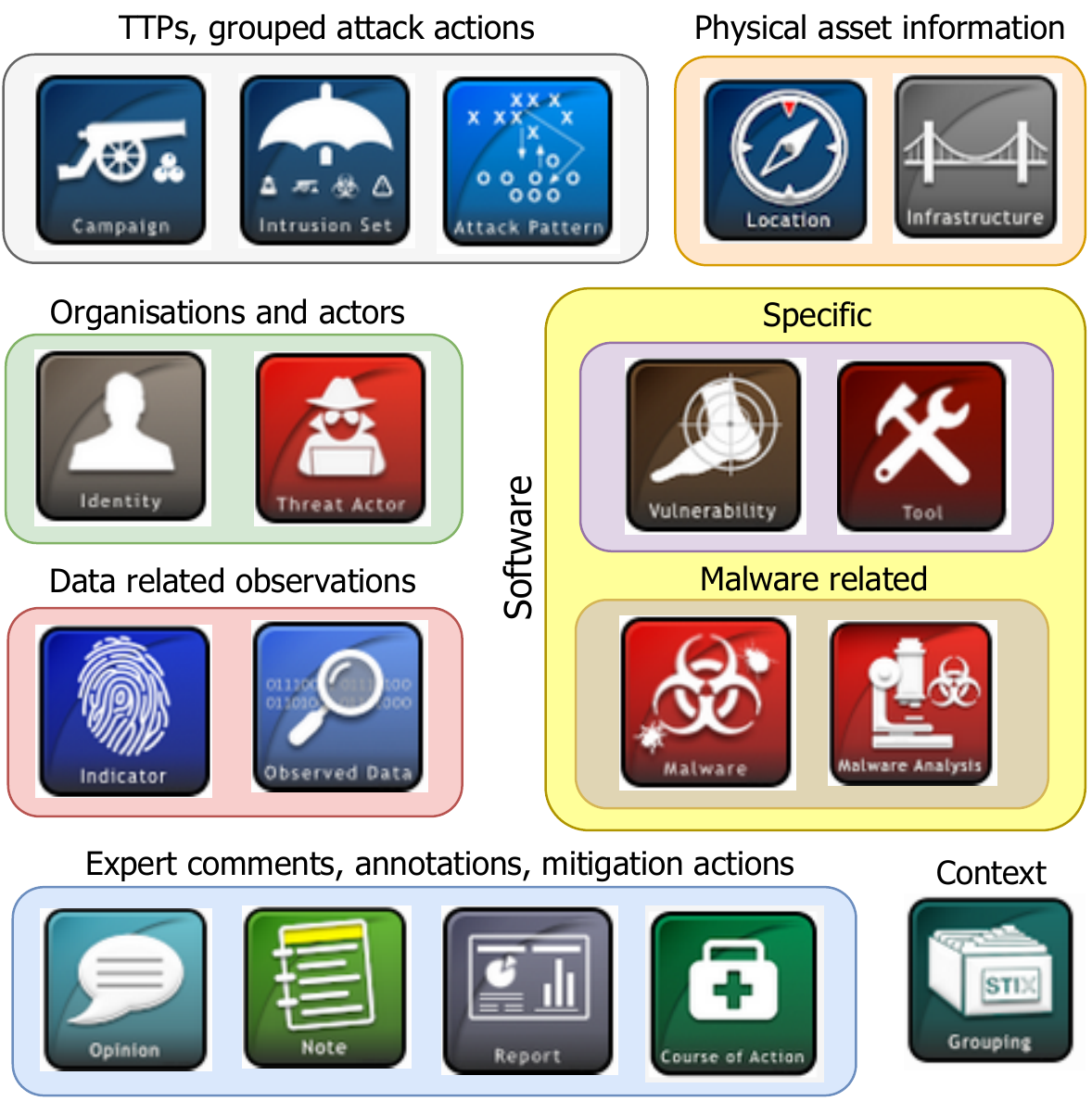}
\caption{Grouping STIX SDOs by characteristic. Based on STIX's Documentation\protect\footnotemark.}
\label{fig:grouping-stix}
\end{figure}

Looking closely at STIX's SDOs and their purposes, one could think about simplifications by grouping its related counterparts.
Figure~\ref{fig:grouping-stix} shows the STIX SDOs grouped by affinity and sharing of concepts.
\footnotetext{\url{https://oasis-open.github.io/cti-documentation/stix/intro}}

Some SDOs are similar and can be grouped together into categories. 
`Attack pattern', `Malware', and `Tool' can all be considered types of TTPs: they describe behaviours and resources that attackers use to carry out their attacks. 
Similarly, `Campaign', `Intrusion set', and `Threat Actor' all describe information about why adversaries carry out attacks and how they organise themselves.

The logical grouping of similar tasks eases modelling and help users focusing on the study's objective.
We mention that analysts should devise STIX models as quickly as cyber-attacks unfold giving a chance for administrators to enact mitigation mechanisms as soon as possible.
We have integrated \toolname{} with the official STIX documentation where, for any element, it provides links to objects and the main description directly on the tool.

\subsection{System design and specification}
Our tool aims to simplify STIX modelling for easier sharing and understanding of both simple and sophisticated malicious incursions in critical infrastructure.
The approach we describe here provides a useful narrative to events as they unfold in the power and telecommunication networks.
The idea is to capture pieces of evidence of malicious incursions, traces, anomalous traffic, or user behaviours and combine with other CTI sources to pro-actively protect the infrastructure against wrongdoings.
We aim at providing clear and direct contextual data of any worth to report inconsistencies happening on any CPS.

\begin{figure*}[!htb]
\centering
\includegraphics[width=0.8\textwidth]{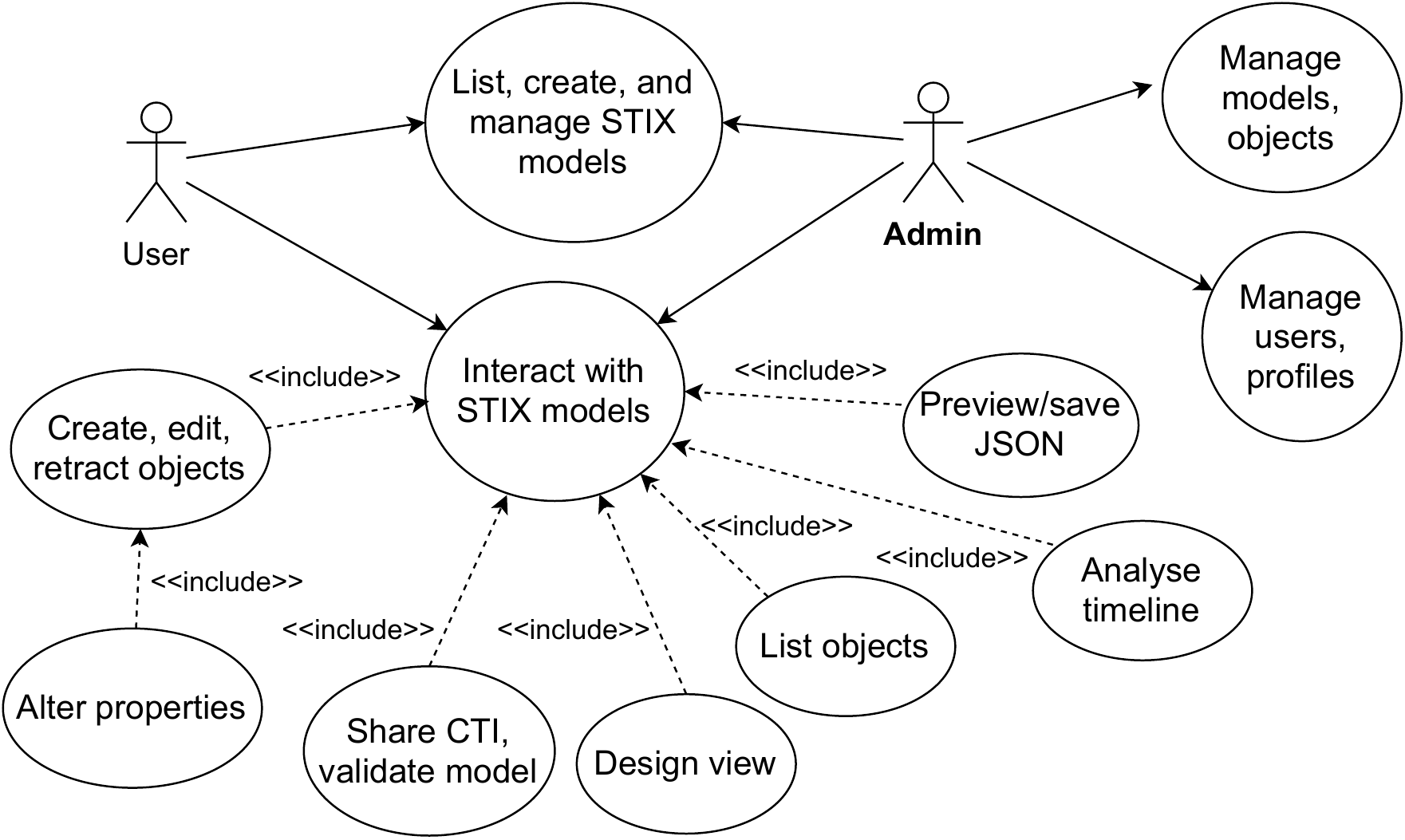}
\caption{Use Case Diagram for \toolname{} showing major functionalities for its existing profiles.}
\label{fig:use-case}
\end{figure*}

Figure~\ref{fig:use-case} depicts the Use Case Diagram with two types of profiles (`User' and `Administrator') and the functionalities they may access after successfully logging in the platform.

The `Preview JSON' function will iterate over the objects of a model and generate excerpts consisting of parameters following the STIX specification.
The list of objects will form a `bundle' that represents the STIX model itself.
The language recommends producing a JavaScript Object Notation (JSON)\footnote{JSON is an open interchange format, offering a structured and prone for validation data structure that is simple to read by both humans and machines.} file containing the model, where the user may aggregate all objects into one definition using the modifier `bundle' (that also has parameters to fill in, such as `id', for instance).
There is also the possibility of downloading the JSON file for any model.

Our tables store the models and objects created by users, that may be normal or administrators.
We are keeping track of when models and objects were created and also modified as well as allowing users to retract (we refrained the word `revoke' because STIX already uses it) objects in models.
Users belong to a profile, and they can see and edit objects created by other users belonging to the same profile.

\subsection{User interface and extended list of features}
We implemented a standard web-based system with so called responsive browser window elements, i.e., flexible hypertext elements that adapts the interface it to fit in different screens (tablets, PC desktops, and even cell phones) and browsers (Opera, IE, Safari, Firefox, or Edge).
For this we have used the framework provided by W3-CSS (from the World-Wide Web consortium, using cascading style sheets)\footnote{\url{https://www.w3schools.com/w3css/default.asp}.}.
The advantage of using W3-CSS over other similar alternatives is due to its simplicity and lack of jQuery/JavaScript elements that were not required by our application.

In the testing environment, we installed a WAMP (Web-Apache-MySQL-PHP) server with standard configuration.
The application in the production server follows the same configuration as the local environment.

We equipped \toolname{} with the following features:
\begin{enumerate}
   \item Simple to use and interact web-based graphical user interface (GUI);
   \item Creation of bespoke STIX models with DBMS persistence, where users (from different profiles) may interact with their definitions; 
   \item Explore the vast parameter space of the STIX specification, identifying types, lists, and likely SDO/SRO/SCO objects to attach to models.
   \begin{itemize}
       \item Assist users when validating models on required/optional parameters, where the system shows types (data structures, lists, strings, integers, or vocabularies), adhering to STIX rules.
   \end{itemize}
   \item Users belonging to the same profile share objects and may edit them as needed that resembles a team analysing a cyber-attack. We have defined five basic profiles: 1) Cyber-security managers; 2) Network administrators; 3) Management; 4) Analysts; and 5) External users.
   \item It allows users to `retract' objects and hidden them from analysis. This is useful for situations where new evidence emerged and the object in that case becomes invalid or retains outdated information.
   \item Validate STIX models (generated in JSON format), where our tool checks required parameters and whether users have chosen valid values as input. Models passing this validation could be used to share CTI across other domains or feeds\footnote{We perform an \textit{internal} validation to check missing parameters, not for checking whether the model is well-formed.}.
   \item Timeline analysis where events are ordered by timestamps parameters, allowing cyber-security officers to consider advanced incursions.
   \begin{itemize}
       \item One could use this feature to quickly update models and differentiate attacks from localised fluctuations in traffic, load, or user activity that could occasionally happen (or any SCO or significant event occurring in the network). Our tool empowers the user to modify models and objects depending on changes in circumstances.
   \end{itemize}
   \item Enable users to `rethink' attacks, updating models' observations, relationships, observable objects, comments, previous analysis.
   \item The system allows STIX model generation (after validation for consistency, e.g., required over optional parameters).
   \item Basic forensic analysis over past events in models and objects.
\end{enumerate}

\toolname{} is implemented as a multi-user web application.
As standard practice, users need to register and log in to the web application to access models.
Models are accessed through identifiers; however, the system would not let an advanced (or malicious) users that guesses identifiers to retrieve a model owned by other users.
We protect the database from SQL injections through \textit{prepared statements} across all queries.
We implement encrypted communications employing HTTP over secure socket layer and, for consistency among peers, we are saving all timestamps in the system with Greenwich Mean Time (GMT).

Besides those features, the tool also provides curated content, with expert domain annotations to establish trustful networks of accredited sources.
It could track cyber-attacks from mere (albeit malicious) observations and sightings until actual manifestation of threats and vulnerability exploitation.
Security officers could use it whilst auditing or performing forensic analysis of cyber-attacks, showing `trails of evidence' that are reviewed throughout a duration, so they understand the timeline.
They could quickly differentiate possible cyber-attacks from normal fluctuations on demand or traffic that occasionally happen in coupled power/telecommunication networks.

Since the tool is directed at end-users sitting on top of the infrastructure, they could use it to better explain cyber-attacks to high-level stakeholders (business managers, and so on) so they could enact changes to prevent further incursions of happening.
Another clear advantage is the use of structured cyber-attack format to ease sharing across similar infrastructure, employing standardised mechanisms used in industry/academy such as STIX (TTP and CKC).
Finally, users may link STIX elements to map previous malicious actions and describe other (more sophisticated) cyber-attacks.
The tool shows cyber-attack progression as it unfolds into networks, matching observations.

\subsection{Showcasing the environment} 
\begin{figure*}[!htb]
\centering
\includegraphics[width=.8\textwidth]{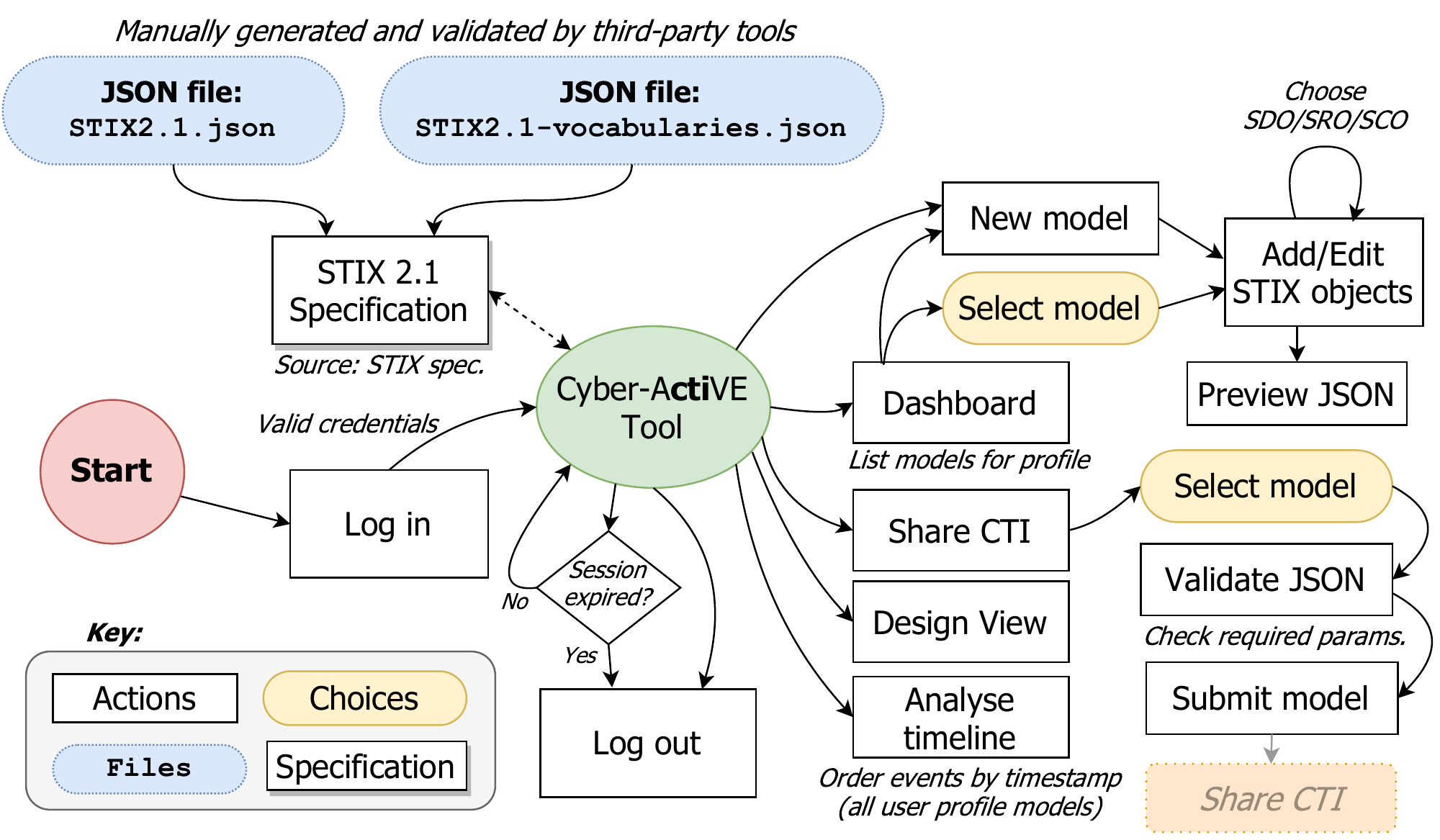}
\caption{Overall process and available features for managing models and objects in \toolname.}
\label{fig:process}
\end{figure*}

\toolname{} employs two manually generated JSON files compiled from the STIX specification: i) a file with basic definitions (\texttt{STIX2.1.json}); and ii) another one for vocabularies (\texttt{STIX2.1-vocabularies.json}).
Those files are public and accessible through the tool.
Figure~\ref{fig:process} shows the process for using these two files to generate input forms containing STIX models and objects.
It covers user's log in whilst operating the system, and then logging out (or forcibly disconnected from its session after a 10-minute timeout), among the rest of the available features.

The figure shows the possible features that users may choose after a successful login.
We implemented the tool with a clean, user friendly, and simple interface where cyber-security officers may select new actions or observations or modify previous actions\footnote{Source code and instructions are publicly available at \url{https://github.com/czekster/cyberactive}.}.
As mentioned earlier, the \toolname{} tool uses a responsive web-based interface for representing cyber-attacks.
Upon entering the tool, we show a brief explanation of what it does and links to its functionalities, such as `My dashboard' that lists all previously created models for that profile.
Users may interact with these elements as they wish as well as inserting and modifying new objects.

\begin{figure}[!htb]
\centering
\fbox{\includegraphics[width=\columnwidth]{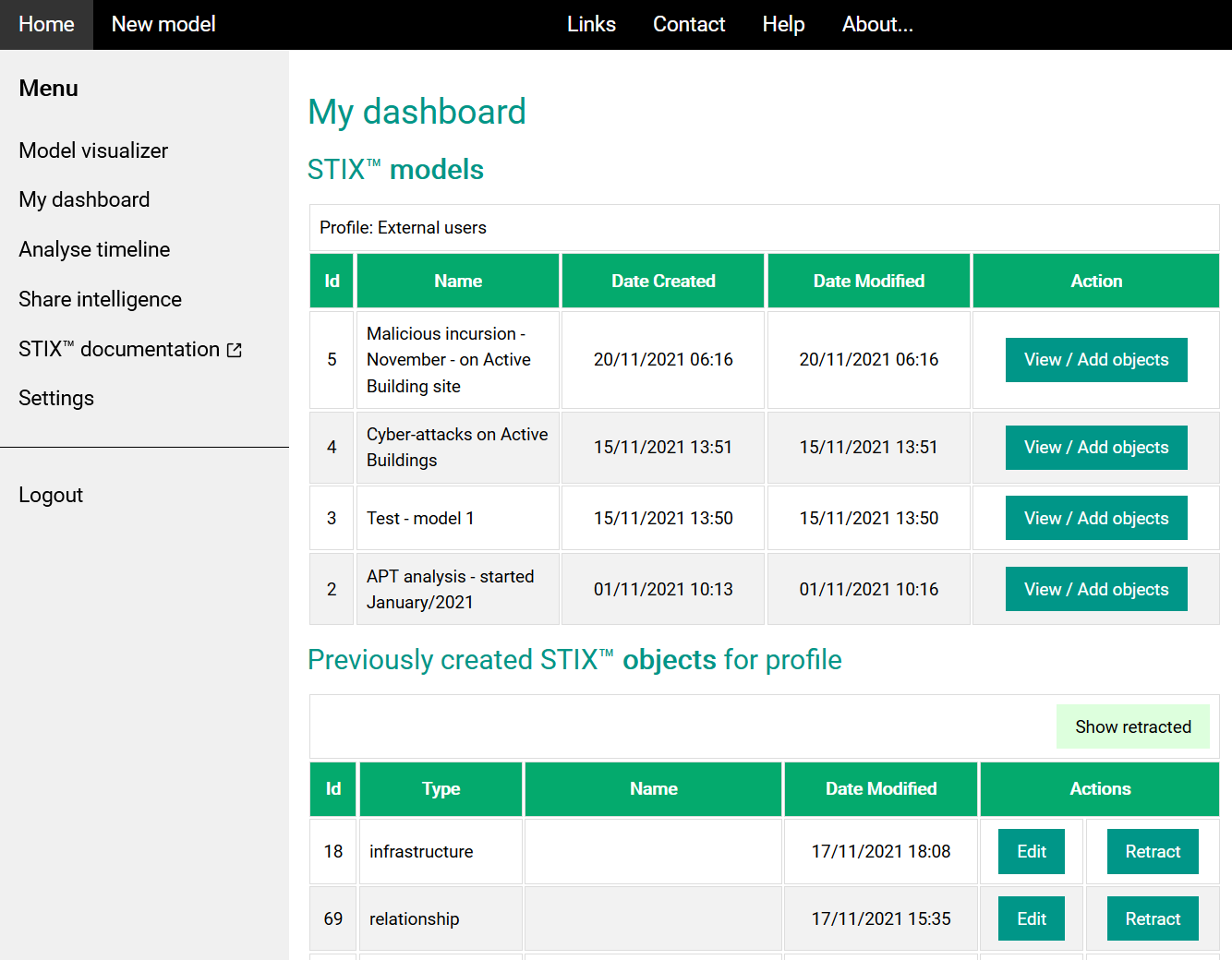}}
\caption{Dashboard feature showing the models and objects created by modellers under the same user profile.}
\label{fig:cap:dashboard}
\end{figure}

We show in Figure~\ref{fig:cap:dashboard} the dashboard for a user, where his/her profile (`Cyber-security managers') has two models.
It is possible, for each model, to 'View/Add objects', and for each object, users may `Edit' or `Retract' (making it temporarily unavailable for using in models -- this could be changed as desired on a link `Restore').
Note that we are ordering models and objects by date of modification (descending).
The platform will automatically paginate the list of models and objects as needed (each page may withhold $10$ instances -- these values could be customised).

\begin{figure}
\begin{subfigure}{0.5\textwidth}
    \centering
    \fbox{\includegraphics[width=.96\linewidth]{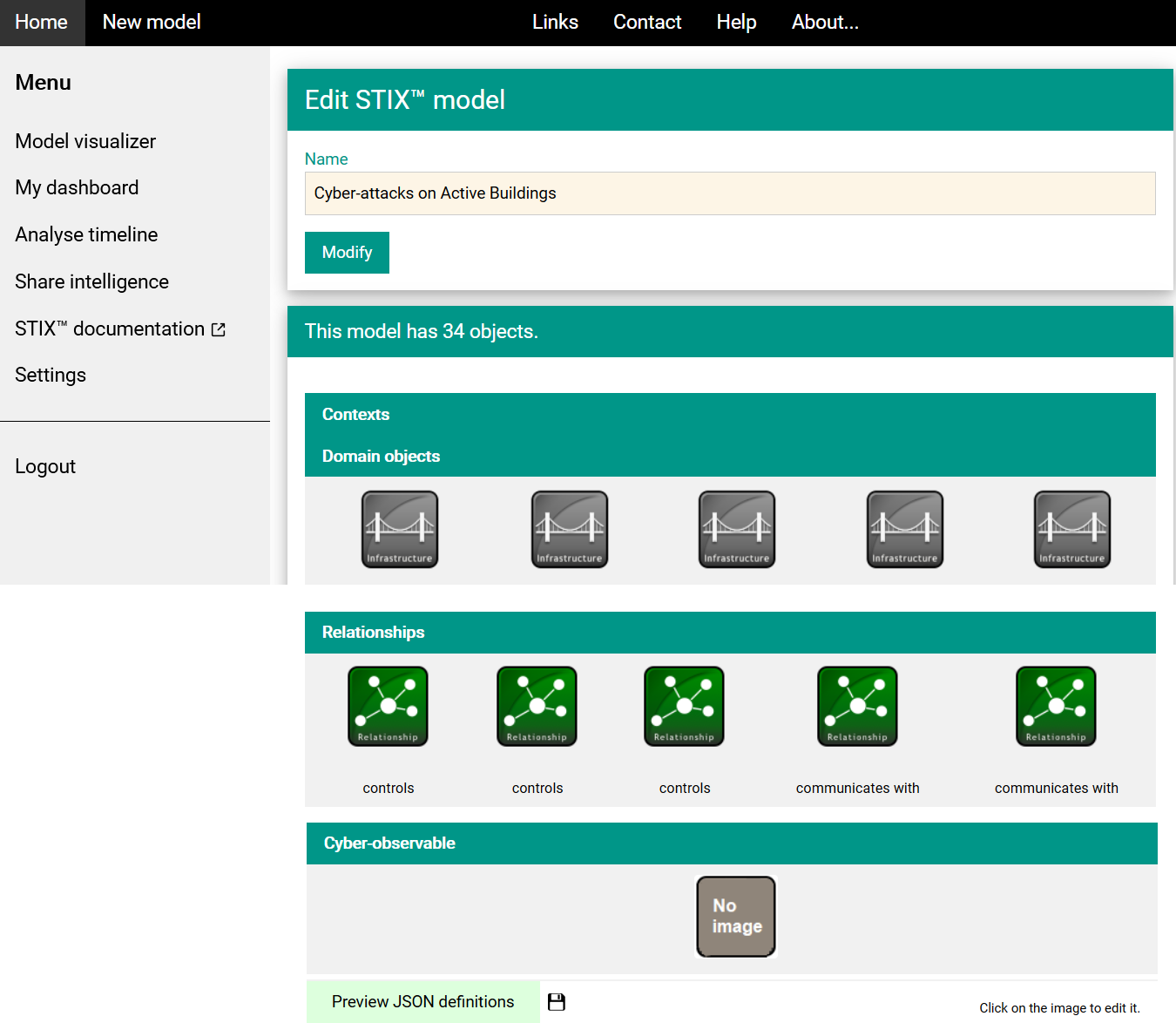}}
    \caption{For editing the model, users may change its `Name' or edit one object that was previously added. The figure was edited and shows only a few objects belonging to this model. On the tool, it lists all the objects.}
    \label{fig:sub-fig-edit-1}
  \end{subfigure}
\begin{subfigure}{0.5\textwidth}
\fbox{\includegraphics[width=.96\linewidth]{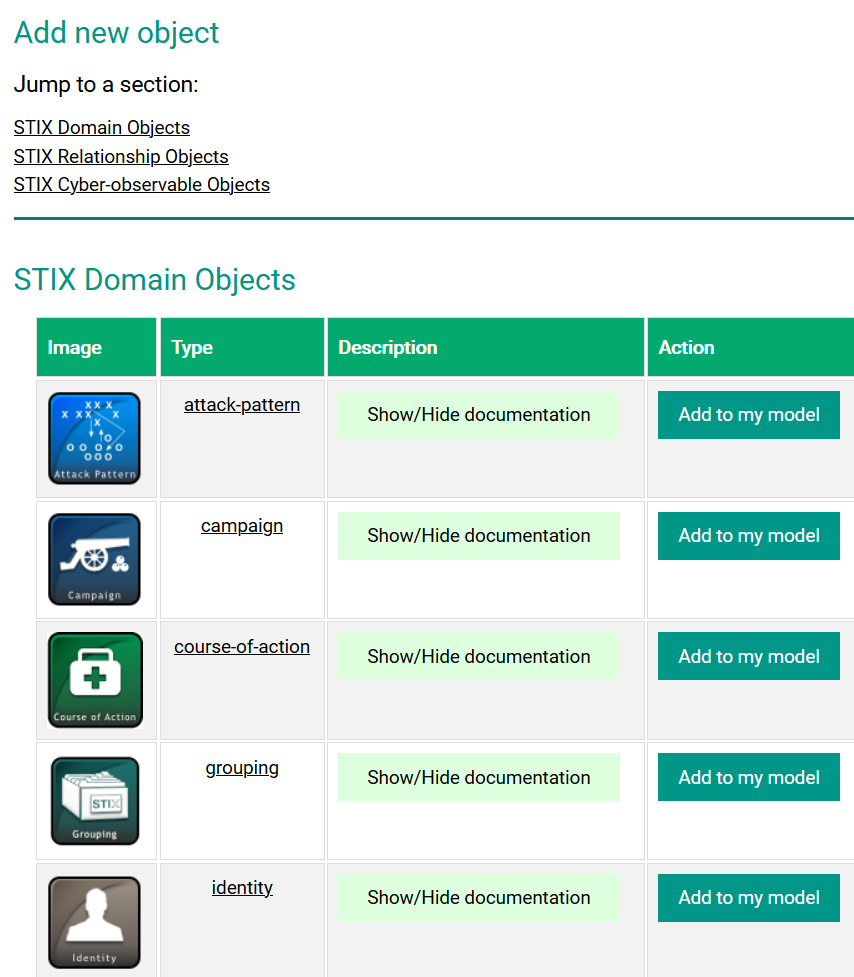}}
    \caption{Users may also add a new STIX object to this model by choosing one alternative from the list and clicking on `Add to my model'.}
    \label{fig:sub-fig-edit-2}
  \end{subfigure}
\caption{Edit STIX model and add objects, where in (\subref{fig:sub-fig-edit-1}) the system shows the current objects that exist for this model and in (\subref{fig:sub-fig-edit-2}) it shows the possibility of adding new objects.}
\label{fig:cap:edit-model}
\end{figure}

Figure~\ref{fig:cap:edit-model} shows the process for editing a model.
The environment shows the model's name (where they could modify it) and the list of previously assigned objects (for the case of the figure, the user has created $11$ objects onto this model).
If the user clicks on any STIX thumbnail the system automatically retrieves the parameters for this object and let the user change it.
Users may preview the current JSON model description by clicking on the button `Preview JSON definitions'.

On the `Edit model' page, users may also add new STIX elements (SDOs, SROS, and SCOs).
The tool will list the thumbnail, the type, the description (that was fetched from the specification).
When the user clicks on `Add to my model', the system will retrieve all properties and let them modify as needed.

\begin{figure}[!htb]
\centering
\fbox{\includegraphics[width=1\columnwidth]{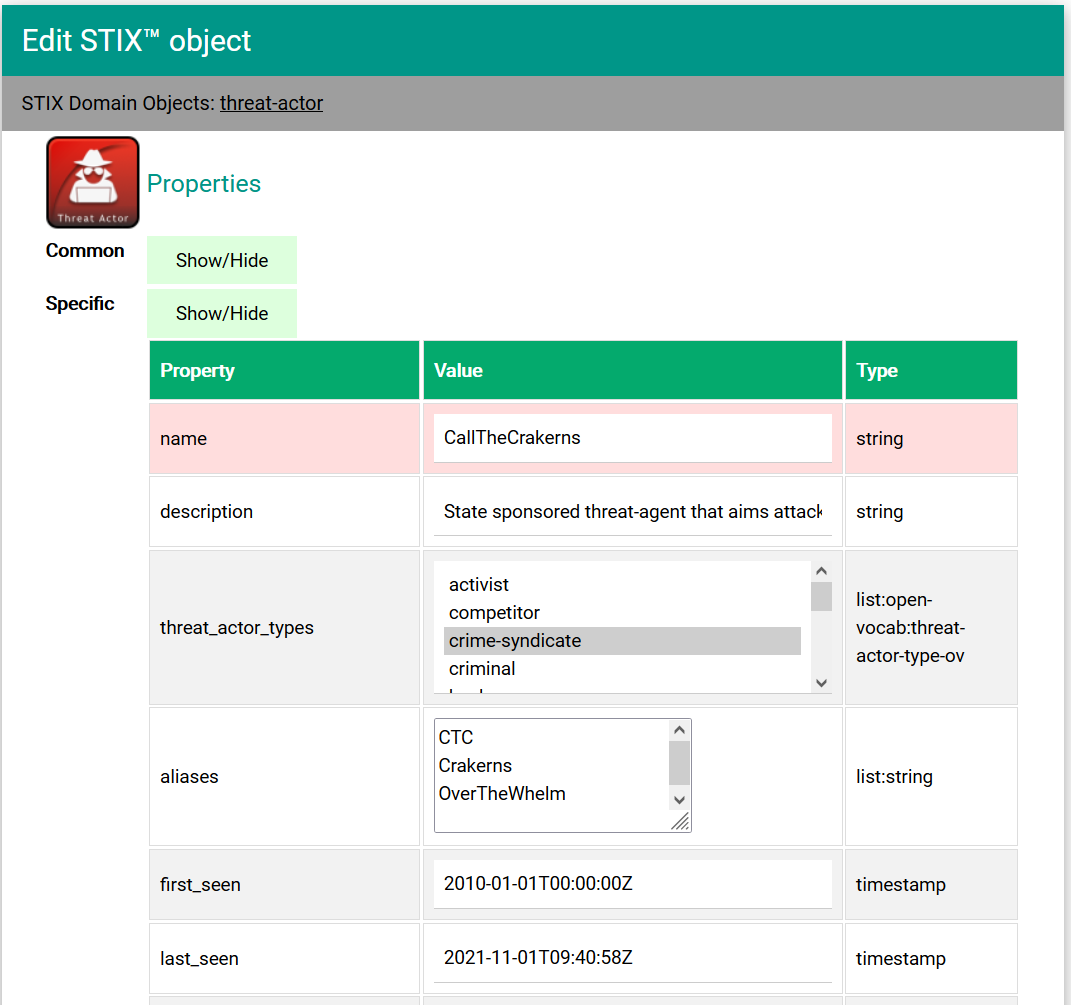}}
\caption{After a user clicks on a STIX thumbnail, the system shows the list of parameters for this object.}
\label{fig:cap:edit-object}
\end{figure}

Users may edit individual STIX objects that are assigned to a specific model.
Figure~\ref{fig:cap:edit-object} shows an excerpt of the parameters for a SDO `Threat actor' that he/she has previously added.
The system shows `Common properties' and `Specific properties', where the buttons toggle Hiding/Viewing screens (to ease navigation).
The system will read the parameters that it has stored on the DBMS and users may change any detail required by simply changing the value on the input fields.

After completing the necessary changes, users may click on the `Submit' button and the system will go back to the model, where they can preview the JSON accordingly.
The system does not support removing objects from models, only retracting (which can be done in the `Dashboard').

\begin{figure}[!htb]
\centering
\fbox{\includegraphics[width=1\columnwidth]{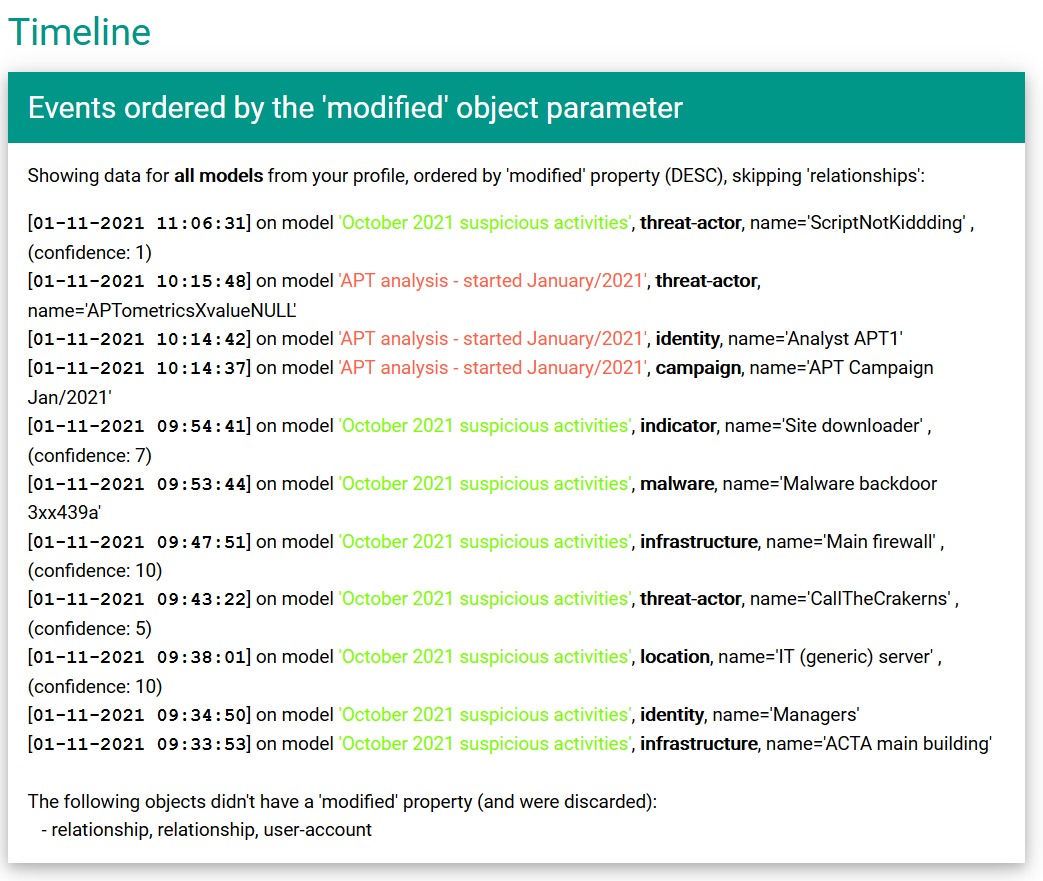}}
\caption{\toolname{} and the `Timeline' feature, showing all modifications ordered by \texttt{modified\_time} DBMS column (if any). The system assigns different colours for all models belonging to the user profile.}
\label{fig:cap:timeline}
\end{figure}

Figure~\ref{fig:cap:timeline} shows the `Timeline' feature, where the system shows events from all models (from the same user profile), ordered by \texttt{modified\_time} property (present in all SDO -- if not present, we show the list of objects without it, in the end).
This feature shows the reasoning and updates behind the process of modelling, showing all the objects users have created and ordering them according to the modification date, by any reason or changes in circumstances.

The system will employ a different colour for each model and will list all objects that users created or modified in a temporal perspective.
This is crucial to understand how cyber-attack unfolded, helping to address mitigations.

\begin{figure}
\begin{subfigure}{0.45\textwidth}
    \centering
    \fbox{\includegraphics[width=.9\linewidth]{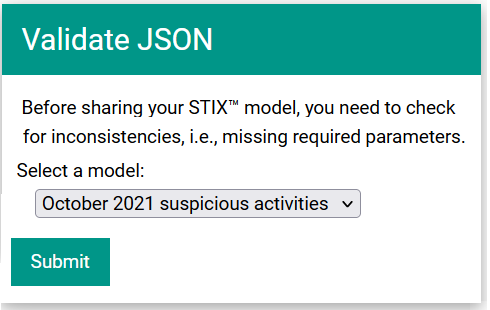}}
    \caption{User selects a model for validation.}
    \label{fig:sub-fig-share-cti-1}
  \end{subfigure}
\begin{subfigure}{0.45\textwidth}
    \centering
    \fbox{\includegraphics[width=.6\linewidth]{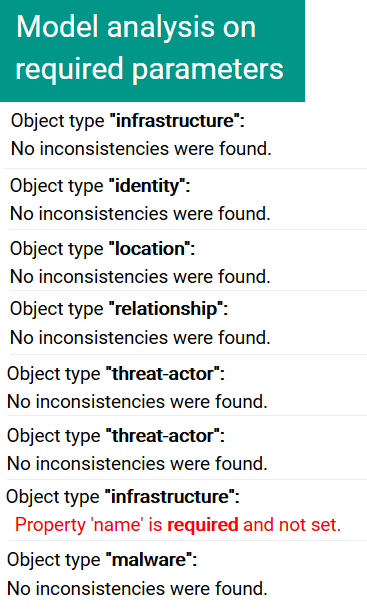}}
    \caption{After selecting the model, the system will check all the objects for required parameters showing a report.}
    \label{fig:sub-fig-share-cti-2}
  \end{subfigure}
\caption{The `Share CTI' allows users to inspect models (and objects) before sharing with other analysts, domain experts, and researchers. After selecting a model (\subref{fig:sub-fig-share-cti-1}), the system validates the model's object by checking the parameters (\subref{fig:sub-fig-share-cti-2}).}
\label{fig:cap:share-cti}
\end{figure}

Finally, we show in Figure~\ref{fig:cap:share-cti} the `Share intelligence' feature.
At the current version of \toolname, it will only verify missing required parameters and point them out to users.

This concludes the major screens and feature description for \toolname.
As stated, it acts as a visual front-end for STIX, where modellers may inspect parameters and assign objects to models for later analysis.

\subsection{Discussion}
The \toolname{} tool aims to enrich CTI-related analysis using more simple, precise, unambiguous, and shareable structured cyber-attack modelling using STIX, a format that is sanctioned by industry and academia.
Cyber-security stakeholders sitting on top of large attack surfaces may profit from CTI to pro-actively set forth measures to thwart malicious actions as they unfold.
Helping users tackle high-level modelling may promote quicker responses and build trustful networks of professionals interested in deterring cyber-criminals in both virtual and physical spaces.

Present work compiled previous research on modelling CTI using STIX and developed a tool to enrich cyber-attack reasoning, guide precise responses, and enhance overall preparedness.
Our tool is based on the standard proposed by STIX and allows users to refine their understanding on detected abnormalities that could potentially lead towards sophisticated attacks.
The features offered by the tool works in both producing STIX models for sharing but also to consume and modify previous ones.
We were motivated to implement such characteristics because simple observations, albeit insignificant, could be a small part of a larger attempt to commit criminal offences.

After analysts input their IoCs and STIX models into our platform, the curated data could help cyber-security officers preparing (before) and mitigating (after) these malicious incursions in a timely fashion.
Our proposition may be highly valuable to identify malicious insiders acting abnormally within the infrastructure.

\section{Conclusions}
\label{sec:conc}

Cyber-security officers working on preventing cyber-attacks in large cyber-physical systems (e.g., a smart building) must provide means to enact effective threat hunting, digital forensics, and CTI.
These measures enormously help managers, analysts, cyber-security officers, and network administrators engage with anomalous behaviours to thwart cyber-attacks.
The integration with `smart' features in sensing or tracking embedded into physical counterparts in the infrastructure will require advanced analysis mechanisms to cope with unusual surges in demand or abnormal happenstances.
In our opinion, CTI plays a crucial role, acting as a useful mechanism to append to other protective mechanisms in place since it provides the context for determining cyber-attacks.
Analysts use threat data feeds from multiple sources to help them understand and respond to malicious incursions.
CTI is still in its early stages as more mature tools and techniques are developed and adopted by organisations.
It must be used in conjunction with other techniques such as attack modelling techniques (AMT)~\cite{lallie2020review} such as attack trees or fault tree analysis, co-simulation~\cite{czekster2021systematic}, focus on APT or LCA, threat modelling, or advanced statistical analysis (ML/AI), to mention a few.

Government is devising incentives for old and new buildings to increase its `smartness' through sensing and remote management features to improve the control over a myriad of distributed assets.
Building managers should consider ways of how to adapt to so called net Zero Energy Building (nZEB) perspectives and enact ways to reduce carbon emissions to meet greener commitments outlined by legislation.
They will push for change in the private and public sector by for instance promoting incentives for customers to purchase equipment and operate as active prosumers in the grid.
So, broader prosumer engagement, dynamic energy pricing and market considerations, utilities, smart settings, and remote-control capabilities will demand thorough cyber-security concerns across the infrastructure.
In this sense, nZEB will become an overspread reality given its advantages.
Beyond helping the climate and ease the strain on power grid on critical hours of the day, ``behind-the-meter'' generation and intelligent storage and release mechanisms will promote energy sharing in the grid network, compensating customers accordingly.

In the ever-changing threat landscape and the ubiquitous use of cloud-based architectures in the smart city, and almost any CPS with IoT, a few measures should be taken into account such as:
\begin{itemize}
    \item \textit{Sharing issues}: organisations have reasons for not sharing CTI: privacy, confidentiality, data related issues and protection. There are clear advantages on sharing, however, industry and academia must discuss advantages and propose new ways of promoting it, through incentives or showing that protective measures do enhance overall cyber-security.
    \item \textit{Update obsolescence}: as the cyber-attack unfolds and gets reported, new venues are explored by adversaries, so older reporting may become outdated.
    \item \textit{Timeliness}: offer updated IoC given emergence of new threats and highly sophisticated cyber-attacks.
    \item \textit{Structured formats}: there is a need for standardised ways of communicating threats, vulnerabilities, and attacks, also on simplified reporting when depicting and learning about malicious incursions.
    \item \textit{Trustfulness}: peers exchanging newest attacks in standardised fashion.
    \item \textit{Model management}: cyber-security officers already have a lot of work deterring cyber-attackers, and modelling should not hinder their activities or impact their productivity. Instead, it should help them and guide better analysis and quick responses.
    \item \textit{Cognitive load}: the magnitude and breadth of data available for analysts could act as the cause for impairing better judgements, given the number of new variables to consider. CTI should offer a minimum set of data points so stakeholders are not overwhelmed by it.
    \item \textit{Scalability}: concerns on emergence of new devices in the infrastructure and reporting.
\end{itemize}

\subsection{Future work and outlook}
Modelling efforts cannot hinder the reasoning or the ability of addressing cyber-attacks quickly, just to strictly follow the standard.
Analysts should be able to describe odd circumstances with as little information as they have at that moment, and only care about modelling details and its constraints afterwards.
In early indications of potential malicious incursions, very little is known about the attacks.
As they unfold and systems gather and compile more evidence, analysts may append and curate preexisting models with this data combined with exterior data sources for full contexts.

We implemented here a front-end editor for STIX modelling, where users may interact with the parameters required by the set of SCO, SRO, or SCO.
The idea is to enrich analysis and allow users to perceive the expected requirements for devising more shareable models to broader audiences.
Timeliness plays a factor in cyber-attacks because one should be able to share possible exposure and vulnerabilities with your trusted peers as soon as possible.

We envision adding more features in future versions of the tool such as integrating \toolname{} with ATT\&CK framework's TTPs and Matrices (provide a static instance of some STIX elements inspired by it, such as existing Threat Actors or mitigations).
We could also accommodate features for analysts such as time-based analysis and to devise ways of tracking the `life-time' of families of cyber-attacks (those focusing on specific assets) and also improving the `Model visualizer' feature.
For instance, we could allow the selection of groups of assets (e.g., all RER, or all IS) and then creating empty objects that will be filled out in another moment.
Because we expect analysts to deploy our tool in their workspaces, we will need to review our features proposition with input from power-based domain experts.
In this evaluation they may specify new streams to look at that are considered essential when inspecting cyber-attacks.
Also, we shall conduct usability testing subjecting users to the tool and inspecting learning curves, whether expectations were met, and incorporating suggestions to improve the tool.

As new implementation, we will consider: i) Increase basic security by logging actions and movements of users, versioning models and objects, showing users older versions visually (e.g., font colour fading); ii) Improve the `timeline' feature and implement the `sharing' CTI feature using actual TAXII servers; iii) Offer to `redact' models and objects before sharing, avoiding unintended disclosure of sensitive data; iv) Implement remaining STIX parameters not tackled by the current tool version, e.g., \texttt{cyber-kill-chain}, \texttt{marking definitions}, and \texttt{dictionary}; v) Force users to provide well-formed input for specific types in accordance with the STIX specification when creating URLs, e-mail addresses, informing (existing/valid) cities or countries; vi) Reuse objects from other previously created models; vii) Allow analysts to operate in different capacities (consulting, analyst, or administrator), across organisations, where they could share infrastructure details and locations; and viii) Ability of exporting and importing models to and from the tool.

There are advantages for implementing systems and employing JSON files to map all objects, types, and vocabularies within the same solution.
Now, any changes in the STIX specification will translate to changes in the JSON files and the system will retain its basic functionality.
Our proposed tool offers interesting features for cyber-security analysis when modelling any malicious incursions in networks.
It makes easier to understand required/optional parameters to enrich models and analysis, besides the ability of sharing models.
Our tool has the potential of easing analysis and capture relevant cyber-security incident data combined with other CTI data sources when documenting and analysing most likely attacks in CI.

Looking at applications, smart buildings pose special concerns to stakeholders addressing cyber-security in power, telecommunications, and building management, to mention a few.
However, older building managers and customers will observe the gains of changing towards smart propositions.
The retrofitting task of converting buildings into smart buildings counterparts will present new challenges for protecting and securing customers participating the network.

\noindent\textbf{Acknowledgements.}
This research was funded by the Industrial Strategy Challenge Fund and EPSRC, EP/V012053/1, Active Building Centre Research Programme (ABC RP).

\bibliographystyle{ieeetr}
\bibliography{biblio}

\appendix

\section{Vulnerability repositories}\label{app:vuln-repos}
Table~\ref{tab:repos} shows a list of vulnerability repositories fed by security officers, organisations, experts, and academia.

\begin{table}[!htb]
\centering
\caption{Vulnerability repositories, TTP, \& scoring systems.}
\resizebox{\columnwidth}{!}{\begin{tabular}{|c|l|l|}
\hline
\textit{Repository} & \multicolumn{1}{c|}{\textit{Description}} & \multicolumn{1}{c|}{\textit{Link}} \\ \hline\hline
NVD & National Vulnerability Database (NIST/US) & \url{https://nvd.nist.gov/} \\ \hline
CVE & Common Vulnerabilities and Exposures & \url{https://cve.mitre.org/} \\ \hline
CWE & Common Weakness Enumeration & \url{https://cwe.mitre.org/} \\ \hline
-- & Metasploit Framework & \url{https://www.metasploit.com} \\ \hline
ATT\&CK & MITRE's ATT\&CK\textsuperscript{TM} Framework & \url{https://attack.mitre.org/} \\ \hline
\textit{attackics} & MITRE's ATT\&CK\textsuperscript{TM} Framework for ICS & Link\tablefootnote{Link: \url{https://collaborate.mitre.org/attackics/index.php}} \\ \hline
exploit-db & Exploit Database & \url{https://www.exploit-db.com/} \\ \hline
rapid7 & rapid7 Vulnerability and Exploit Database & \url{https://www.rapid7.com/db/} \\ \hline
CPE & Common Platform Enumeration & \url{http://cpe.mitre.org/} \\ \hline
-- & CERT/CC vulnerability notes & \url{https://kb.cert.org/vuls/} \\ \hline
ZDI & TrendMicro’s Zero Day Initiative & Link\tablefootnote{Link: \url{https://www.zerodayinitiative.com/advisories/published/}} \\ \hline
\multicolumn{3}{l}{\textit{Scoring system}} \\ \hline
CVSS & Common Vulnerabilities Scoring system & \url{https://www.first.org/cvss/} \\ \hline
\end{tabular}
}
\label{tab:repos}
\end{table}

\end{document}